\documentclass[a4paper,11pt]{article}
\pdfoutput=1 
\usepackage{jheppub}
\usepackage[T1]{fontenc} 
\usepackage{color}
\usepackage{fancybox}
\usepackage{amsmath}
\usepackage{amsfonts}
\usepackage{slashed}
\usepackage{amssymb}
\usepackage{indentfirst}
\usepackage{stmaryrd}
\usepackage{subcaption}
\usepackage{graphicx}
\usepackage{cleveref}
\usepackage{textcomp}
\usepackage[toc]{appendix}
\usepackage[font={small,it}]{caption}
\usepackage[justification=centering]{caption}

\allowdisplaybreaks

\newcommand{\be}{\begin{equation}}
\newcommand{\bea}{\begin{eqnarray}}

\newcommand{\ee}{\end{equation}}
\newcommand{\eea}{\end{eqnarray}}

\title{\boldmath Notes on the 11D pure spinor wordline vertex operators}

\author{Max Guillen$^{*}$}

\affiliation{$^{*}$Department of Physics and Astronomy, Uppsala University, 75108 Uppsala, Sweden
}

\emailAdd{max.guillen@physics.uu.se}

\abstract{The construction of the ghost number zero and one vertex operators for the 11D pure spinor superparticle will be revisited. In this sense, an alternative way of defining the ghost number one vertex operator will be given after introducing a ghost number -2 operator made out of physical operators defined on the 11D non-minimal pure spinor superspace. This procedure will make explicit and transparent the relation between the ghost number three and one vertex operators. In addition, using a non-Lorentz covariant b-ghost, ghost number zero and two vertex operators satisfying standard descent equations will be presented in full form.    
}


\begin{document} 
\hfill{}
\maketitle

\section{Introduction}
11D supergravity \cite{Cremmer:1978km} has been conjectured to be the low-enery limit of the so-called M-Theory \cite{Hull:1994ys,Witten:1995ex}. Since 11D supergravity naturally couples to a supermembrane, this latter was proposed to be the underlying worldvolume theory whose physical spectrum would be the UV-completion of 11D supergravity \cite{Duff:1987bx,Bergshoeff:1987cm}. The importance of its study relies on the direct relationship between the eleventh coordinate length and the closed string coupling constant, which implies that M-Theory interactions encode non-perturbative information of string theory.

\vspace{2mm} 
On the other hand, the pure spinor formalism for superstrings \cite{Berkovits:2000fe} has been shown to be extremely powerful and efficient for scattering amplitudes computations as compared to the traditional RNS and Green-Schwarz superstrings. This nice feature, among others, is a direct consequence of the built-in manifest supersymmetry, the simplicity of its BRST operator, and the properties of 10D pure spinor variables. Using similar ideas, Berkovits proposed in \cite{Berkovits:2002uc} a pure spinor formulation for studying supermembranes. As for superstrings, this pure spinor supermembrane exhibits manifest super-Poincar\'e symmetry, and its BRST operator is given by a simple expression involving the fermionic Green-Schwarz supermembrane constraint. However, unlike the 10D case, the 11D pure spinor variables are required to satisfy extra constraints in such a way that the supermebrane action and BRST operator are BRST-invariant and nilpotent, respectively.

\vspace{2mm}
A particular limit of the pure spinor supermembrane, namely the one obtained when all the worldvolume fields are independent of the spatial worldvolume coordinates, defines the so-called 11D pure spinor superparticle. The light-cone Hilbert spaces corresponding to this pure spinor worldline model and the 11D Brink-Schwarz-like superparticle \cite{Green:1999by} have been shown to be equivalent to each other \cite{Guillen:2017mte}. More covariantly, this pure spinor superparticle model can actually be shown to describe the 11D supergravity degrees of freedom in its Batalin-Vilkovisky description, where physical states are located at ghost number three cohomology \cite{Berkovits:2002uc}. In this manner, 11D pure spinor worldline correlators might provide us an efficient way to calculate 11D supergravity scattering amplitudes in a manifestly supersymmetric way. For such a purpose, an 11D N-point correlation function prescription was given in \cite{Anguelova:2004pg}. This involves two ghost number three vertex operators, one ghost number one vertex operator and N-3 ghost number zero vertex operators, so that the seven zero modes of the 11D pure spinor variable are completely saturated. Thus, the existence and construction of the ghost number one and zero vertex operators play a crucial role from a phenomenological viewpoint.

\vspace{2mm}
A BRST-closed ghost number one worldline vertex operator has been recently constructed in \cite{Berkovits:2019szu}. The 3-point correlation function involving this operator was shown to match the 3-point coupling of the 11D supergravity pure spinor master action \cite{Guillen:2020zwm,Cederwall:2010tn}. However, as discussed before, higher-point correlators require the construction of the ghost number zero vertex. This problem was also tackled in \cite{Berkovits:2019szu}, and it was reported an incompatibility between the existence of this operator in the minimal formalism, satisfying a standard descent equation, and 11D supergravity. Nevertheless, an alternative solution using the non-minimal 11D b-ghost \cite{Cederwall:2012es,Berkovits:2017xst} was proposed, even though no explicit expressions for it was calculated.

\vspace{2mm}
In this paper, we revisit the construcion of the ghost number one and zero vertex operators. Concretely, we show the ghost number one vertex operator can actually be obtained from the ghost number three vertex operator via a ghost number -2 operator made out of the so-called physical operators. These physical operators were introduced in \cite{Cederwall:2009ez,Cederwall:2011vy} for constructing pure spinor master actions for maximally supersymmetric gauge theories \cite{Cederwall:2010tn,Cederwall:2013vba}. They describe non-trivial cohomology up to shift symmetry terms, which can be read off from superspace equations of motion. In this way, the relationship between the ghost number three and one operators is made precise and transparent. In addition to this, we construct a ghost number zero vertex operator using a non-Lorentz covariant or Y-formalism-like 11D b-ghost acting on the ghost number one vertex operator. Although this b-ghost is not Lorentz-covariant, its variation under Lorentz-transformations is BRST-exact, so that the ghost number zero vertex operator is Lorentz-invariant up to BRST-exact terms. By construction, this ghost number zero vertex will also be invariant under local gauge transformations up to BRST-exact and surface terms, and satisfy a standard descent equation. However, unlike the 10D case, this vertex will present explicit dependence on the non-Lorentz covariant spinor $Y_{\alpha}$ which makes difficult its use for practical purposes. We elaborate a discussion on this point and present some thoughts which might eventually help us to fix this problem. For completeness, we also derive a ghost number two vertex operator which is related to the ghost number three vertex through a standard descent equation.

\vspace{2mm}
The paper is organized as follows. We review the 11D pure spinor superparticle and its quantization in section \ref{section2}. 11D physical operators are introduced in section \ref{section3}, and an explicit map constructed out of them relating the ghost number three and one vertex operators is presented. In section \ref{section4}, the ghost number zero vertex operator is explicitly constructed by making use of a non-Lorentz covariant b-ghost. In order to illustrate this procedure in a simpler way, we give a brief review on its 10D analogue in Appendix \ref{appendixA}
. In section \ref{section5}, we apply similar ideas to the ghost number three vertex and define a ghost number two vertex operator. Finally, we discuss our results in section \ref{section6}.

\section{11D pure spinor superparticle}\label{section2}
The eleven-dimensional pure spinor superparticle action is given by \cite{Berkovits:2002uc,Guillen:2017mte}
\begin{eqnarray}\label{11dpssaction}
S &=& \int d\tau \left[P_{m}\partial_{\tau}X^{m} + p_{\mu}\partial_{\tau}\theta^{\mu} + w_{\alpha}(\partial_{\tau}\lambda^{\alpha}+ \partial_\tau Z^M
\Omega_{M\beta}{}^\alpha \lambda^\beta)\right]
\end{eqnarray}
We use Greek/Latin letters from the beginning of the alphabet to denote tangent-space eleven-dimensional spinor/vector indices, and Greek/Latin letters from the middle of the alphabet to denote coordinate-space eleven-dimensional spinor/vector indices. Furthermore, capital letters from the
beginning/middle of the alphabet will denote tangent/coordinate-superspace indices. In \eqref{11dpssaction}, $X^{m}$ is an eleven-dimensional spacetime coordinate, $\theta^{\mu}$ is an eleven-dimensional Majorana spinor, $Z^M=(X^m, \theta^\mu)$, $\lambda^{\alpha}$ is a bosonic eleven-dimensional Majorana spinor satisfying $\lambda\Gamma^{a}\lambda = 0$; $P_{m}$, $p_{\mu}$, $w_{\alpha}$ are the conjugate momenta associated to $X^{m}$, $\theta^{\mu}$, $\lambda^{\alpha}$ respectively, and $\Omega_{M\beta}{}^\alpha$ is the background super-spin-connection. The $32\times 32$ symmetric matrices $(\Gamma^{a})^{\alpha\beta}$ and $(\Gamma^{a})_{\beta\delta}$ are the 11D gamma-matrices which satisfy $(\Gamma^{(a})^{\alpha\beta}(\Gamma^{b)})_{\beta\delta} = \eta^{ab}\delta^{\alpha}_{\delta}$. The BRST operator is given by
\begin{eqnarray}\label{minimalBRSToperator}
Q_{0} &=& \lambda^{\alpha}d_{\alpha}
\end{eqnarray}
where
\begin{eqnarray}
d_\alpha &=& E_{\alpha}^{\hspace{2mm}M} (P_M +\Omega_{M\beta}{}^\gamma w_\gamma \lambda^\beta)
\end{eqnarray}
In a flat Minkowski background, $d_{\alpha} = P_{\alpha} - \frac{1}{2}(\Gamma^{m}\theta)_{\alpha}P_{m}$ are the fermionic constraints of the 11D Brink-Schwarz-like superparticle. The nilpotency of $Q_{0}$ straightforwardly follows from the pureness of $\lambda^{\alpha}$. We have written down a subscript $0$ in \eqref{minimalBRSToperator} to indicate this operator corresponds to the minimal formalism of the 11D superparticle \eqref{11dpssaction}. A slight modification of \eqref{minimalBRSToperator} will take place later on when we introduce non-minimal pure spinor variables. 

\vspace{2mm}
The physical spectrum is defined as the BRST-cohomology of $Q_{0}$. One can show that the eleven-dimensional linearized supergravity physical fields are described by the ghost number three state: $U^{(3)} = \lambda^{\alpha}\lambda^{\beta}\lambda^{\delta}C_{\alpha\beta\delta}$ \cite{Berkovits:2002uc}, where $C_{\alpha\beta\delta}$ is the lowest dimensional component of the 11D supergravity super-three-form $C_{ABC}$. To see this, one writes the most general ghost number three superfield $U^{(3)} = \lambda^{\alpha}\lambda^{\beta}\lambda^{\delta}\tilde{C}_{\alpha\beta\delta}$ and imposes on it the physical state conditions. In this way, one finds that $\tilde{C}_{\alpha\beta\delta}$ must obey the following equations of motion and gauge transformations
\begin{eqnarray}\label{eqeqe1}
D_{(\alpha}\tilde{C}_{\beta\delta\epsilon)} &=& (\Gamma^{a})_{(\alpha\beta}\tilde{C}_{\vert a\vert \delta\epsilon)}\nonumber\\
\delta \tilde{C}_{\alpha\beta\delta} &=& D_{(\alpha}\Lambda_{\beta\delta)}
\end{eqnarray}
for some arbitrary superfield $\Lambda_{\beta\delta}$. These are exactly the superspace constraints describing linearized 11D supergravity \cite{Brink:1980az}, and so one identifies $\tilde{C}_{\alpha\beta\delta} = C_{\alpha\beta\delta}$. Its lowest order theta-expansion reads
\begin{eqnarray}
C_{\alpha\beta\delta} &=& (\Gamma^{a}\theta)_{\alpha}(\Gamma^{b}\theta)_{\beta}(\Gamma^{c}\theta)_{\delta}C_{abc}(x) + (\Gamma^{(a}\theta)_{\alpha}(\Gamma^{b)c}\theta)_{\beta}(\Gamma_{c}\theta)_{\delta}h_{ab}(x)\nonumber\\
&& + (\Gamma^{b}\theta)_{\alpha}[(\Gamma^{c}\theta)_{\beta}(\Gamma^{d}\theta)_{\delta}(\theta\Gamma_{cd})_{\epsilon} - (\Gamma_{cd}\theta)_{\beta}(\Gamma_{c}\theta)_{\delta}(\Gamma_{d}\theta)_{\epsilon}]\chi_{b}^{\epsilon}(x) + \ldots
\end{eqnarray}
where the fields $C_{abc}(x)$, $h_{ab}(x)$, $\chi_{b}^{\alpha}(x)$ satisfy the linearized 11D supergravity equations of motion and gauge invariances
\begin{eqnarray}
\partial^{c}[\partial_{c}h_{ab} - 2\partial_{(a}h_{b)c}] - \partial_{a}\partial_{b}h^{c}{}_{c} = 0\hspace{2mm}&,&\hspace{2mm} \delta h_{ab} = \partial_{(b}\Lambda_{c)}\nonumber\\
\partial^{d}\partial_{[a}C_{bcd]} = 0\hspace{2mm}&,&\hspace{2mm} \delta C_{abc} = \partial_{[a}\Lambda_{bc]}\nonumber\\
(\Gamma^{abc})_{\alpha\beta}\partial_{b}\chi_{c}^{\beta} = 0 \hspace{2mm}&,&\hspace{2mm} \delta \chi_{a}^{\beta} = \partial_{a}\Lambda^{\beta}
\end{eqnarray}
where $\Lambda_{a}$, $\Lambda_{bc}$, $\Lambda^{\alpha}$ are arbitrary gauge parameters. Using the reducibility or fixed point methods, it can be shown that the remaining non-trivial cohomology is found at ghost number 0, 1, 2, 4, 5, 6 and 7 states; describing the ghosts, antifields and antighosts of linearized 11D supergravity in its BV formulation.

\vspace{2mm}
Using the 11D scalar top cohomology, one can define a BRST-invariant and supersymmetric measure as $\langle \lambda^{7}\theta^{9} \rangle = 1$ for computing scattering amplitudes of N external states. This measure has been successfully used to get the kinetic terms of the 11D supergravity action from a second-quantized point of view \cite{Berkovits:2002uc}. The tree-level N-point amplitude is then given by the N-point correlator \cite{Anguelova:2004pg}
\begin{eqnarray}\label{11amplitudeprescription}
\mathcal{A}^{11D}_{N} &=& \langle U^{(3)}_{1}(\tau_{1})U^{(3)}_{2}(\tau_{2})U^{(1)}_{3}(\tau_{3})\int d\tau_{4}U^{(0)}_{4}(\tau_{4})\ldots \int d\tau_{N} U^{(0)}_{N}(\tau_{N})\rangle
\end{eqnarray}
In this expression $U^{(3)}$ is the ghost number three superparticle vertex operator introduced above, $U^{(1)}$ is a ghost number one vertex operator, and $U^{(0)}$ is a vertex operator of ghost number zero. The properties these operators must satisfy can easily be obtained from the pure spinor supermembrane formulation in the appropriate limit \cite{Berkovits:2002uc}. Thus, one finds that
\begin{eqnarray}\label{supermembranedescentequations}
\{Q, U^{(3)}\} = 0 \hspace{2mm},\hspace{2mm} \{Q , U^{(2)}\} = \partial_{\tau}U^{(3)} \hspace{2mm},\hspace{2mm}  \{Q, U^{(1)}\} = 0 \hspace{2mm},\hspace{2mm}\{Q, U^{(0)}\} &=& \partial_{\tau}U^{(1)}
\end{eqnarray}
The operator $U^{(1)}$ has been recently constructed in \cite{Berkovits:2019szu}. We briefly review this construction in next section, and present a compact relation between it and $U^{(3)}$. On the other hand, the operator $U^{(0)}$ was shown to be incompatible with 11D supergravity in the minimal framework. We address this issue in section \ref{section4} after introducing a non-Lorentz covariant or Y-formalism-like 11D b-ghost. Using this very same b-ghost, a ghost number two vertex $U^{(2)}$ is explicitly constructed in section \ref{section5}.

\vspace{2mm}
An alternative prescription for computing N-point correlators makes use of the ghost number four vertex operator $U^{(4)}$ which contains the antifields of the 11D supergravity physical fields \cite{Berkovits:2002uc}. In this scheme, the ghost number zero operator $U^{(0)}$ plays a crucial role even for the 3-particle amplitude. Therefore, the existence and construction of $U^{(0)}$ is fundamental for the pure spinor wordline model to provide concrete predictions on 11D supergravity interactions.

\vspace{2mm}
Despite the nice features of the 11D pure spinor measure $\langle \lambda^{7}\theta^{9}\rangle = 1$ mentioned above, it is a degenerate quantity, and thus it is not a well-defined integration measure. In order to solve this problem, one introduces non-minimal pure spinor variables \cite{Berkovits:2005bt} consisting of a pair of conjugate variables $(\bar{\lambda}_{\alpha}, \bar{w}^{\beta})$, $(r_{\alpha}, s^{\beta})$ satisfying $\bar{\lambda}\Gamma^{a}\bar{\lambda} = \bar{\lambda}\Gamma^{a}r = 0$, through the quartet argument, which means the BRST operator is modified to
\begin{eqnarray}
Q &=& Q_{0} + r_{\alpha}\bar{w}^{\alpha}
\end{eqnarray}
so that the cohomology remains the same as before. One then defines an integration measure on non-minimal pure spinor superspace as \cite{Cederwall:2009ez}
\begin{eqnarray}
[dZ] &=& d^{11}X\,d^{32}\theta\,[d\lambda][d\bar{\lambda}][dr]\,\mathcal{N}
\end{eqnarray}
where
\begin{eqnarray}
\left[d\lambda\right]\lambda^{\alpha_{1}}\ldots\lambda^{\alpha_{7}} &=& (\epsilon T^{-1})^{\alpha_{1}\ldots\alpha_{7}}_{\hspace{9mm}\beta_{1}\ldots\beta_{23}}d\lambda^{\beta_{1}}\ldots d\lambda^{\beta_{23}}\nonumber\\
\left[d\bar{\lambda}\right]\bar{\lambda}_{\alpha_{1}}\ldots\bar{\lambda}_{\alpha_{7}} &=& (\epsilon T)_{\alpha_{1}\ldots\alpha_{7}}^{\hspace{9mm}\beta_{1}\ldots\beta_{23}}d\bar{\lambda}_{\beta_{1}}\ldots d\bar{\lambda}_{\beta_{23}}\nonumber\\
\left[d r\right] &=& (\epsilon T^{-1})^{\alpha_{1}\ldots\alpha_{7}}_{\hspace{9mm}\beta_{1}\ldots\beta_{23}}\bar{\lambda}_{\alpha_{1}}\ldots\bar{\lambda}_{\alpha_{7}}(\frac{\partial}{\partial r_{\beta_{1}}})\ldots (\frac{\partial}{\partial r_{\beta_{23}}})
\end{eqnarray}
The Lorentz-invariant tensors $(\epsilon T)_{\alpha_{1}\ldots\alpha_{7}}^{\hspace{9mm}\beta_{1}\ldots\beta_{23}}$ and $(\epsilon T^{-1})^{\alpha_{1}\ldots \alpha_{7}}_{\hspace{9mm}\beta_{1}\ldots\beta_{23}}$ were defined in \cite{Cederwall:2009ez}. They are symmetric and gamma-traceless in $(\alpha_{1},\ldots,\alpha_{7})$ and are antisymmetric in $[\beta_{1},\ldots ,\beta_{23}]$. $\mathcal{N}$ is a regularization factor which is given by $\mathcal{N} = e^{-\lambda\bar{\lambda} - r\theta}$. Since the measure converges as $\lambda^{16}\bar{\lambda}^{23}$ when $\lambda\to 0$, the tree-level correlator is well-defined if the integrand in \eqref{11amplitudeprescription} diverges slower than  $\lambda^{-16}\bar{\lambda}^{-23}$.

\section{Ghost number one vertex operator}\label{section3}
The ghost number one vertex operator was constructed in \cite{Berkovits:2019szu} as a first order deformation to the BRST operator $Q_{0} = \lambda^{\alpha}d_{\alpha}$ in a flat background. It takes the explicit form
\begin{eqnarray}\label{definitionofU1}
U^{(1)} &=& \lambda^{\alpha}h_{\alpha}{}^{a}P_{a} - \lambda^{\alpha}h_{\alpha}{}^{\beta}d_{\beta} - \lambda^{\alpha}\Omega_{\alpha\,\beta}{}^{\delta}\lambda^{\beta}w_{\delta}
\end{eqnarray}
where $h_{\alpha}{}^{a}$, $h_{\alpha}{}^{\beta}$ are the deformations of $D_{\alpha}$ along the directions $D_{\beta}$, $\partial_{a}$, and $\Omega_{\alpha\,\beta}{}^{\delta}$ is the lowest-dimensional component of the super-spin-connection $\Omega_{A\,B}{}^{C}$. Using the linearized 11D supergravity equations of motion
\begin{eqnarray}
2 D_{(\alpha}h_{\beta)}{}^{a} + 2 h_{(\alpha}{}^{\delta}(\Gamma^{a})_{\beta)\delta} - h_{b}{}^{a}(\Gamma^{b})_{\alpha\beta} &=& 0\label{eomforU1a}\\
2D_{(\alpha}h_{\beta)}{}^{\delta} - 2\Omega_{(\alpha\beta)}{}^{\delta} - (\Gamma^{a})_{\alpha\beta}h_{a}{}^{\delta} &=& 0\label{eomforU1b}\\
R_{\alpha\beta, c}{}^{d} - 2 D_{(\alpha}\Omega_{\beta)c}{}^{d} + (\Gamma^{a})_{\alpha\beta}\Omega_{a\,c}{}^{d} &=& 0\label{eomforU1c}\\
\frac{1}{4}(\Gamma^{ab})_{(\delta}{}^{\gamma}R_{\alpha\beta),ab} + (\Gamma^{a})_{(\alpha\beta}T_{\delta)a}{}^{\gamma} &=& 0\label{eomforU1d}\\
\partial_{a}h_{\alpha}{}^{b} - D_{\alpha}h_{a}{}^{b} + h_{a}{}^{\beta}(\Gamma^{b})_{\beta\alpha} + \Omega_{\alpha\,a}{}^{b} &=& 0\label{eomforU1g}\\
\partial_{a}h_{\alpha}{}^{\beta} - D_{\alpha}h_{a}{}^{\beta} + T_{a\alpha}{}^{\beta} - \Omega_{a\,\alpha}{}^{\beta} &=& 0\label{eomforU1e}\\
R_{a\alpha,\beta}{}^{\delta} - \partial_{a}\Omega_{\alpha\,\beta}{}^{\delta} +  D_{\alpha}\Omega_{a\,\beta}{}^{\delta} &=& 0\label{eomforU1f}
\end{eqnarray}
one can show that $U^{(1)}$ is BRST-closed. Moreover, it is invariant under super-gauge transformations up to BRST-exact terms.

\vspace{2mm}
Next, we will derive $U^{(1)}$ in \eqref{definitionofU1} from the ghost number three vertex $U^{(3)}$ studied in section \ref{section2} through a ghost number -2 operator constructed out of non-minimal pure spinor variables.
\subsection{Physical operators}
As done in 10D \cite{Cederwall:2011vy}, one can convert eqns. \eqref{eomforU1a}-\eqref{eomforU1d} into operator equations after introducing the so-called physical operators. For this purpose, one multiplies on both sides of \eqref{eomforU1a}-\eqref{eomforU1d} by $\lambda^{\alpha}\lambda^{\beta}$ to get
\begin{eqnarray}
Q(\Phi^{a})+ (\lambda\Gamma^{a})_{\alpha}\Phi^{\alpha} &=& 0\\
Q(\Phi^{\alpha}) - \lambda^{\beta}\Phi_{\beta}{}^{\alpha} &=& 0\\
Q(\Phi_{\alpha}{}^{\beta}) + (\lambda\gamma^{a})_{\alpha}\mathcal{T}_{a}{}^{\beta} &=& 0
\end{eqnarray}
where $\Phi^{a} = \lambda^{\alpha}h_{\alpha}{}^{a}$, $\Phi^{\alpha} = \lambda^{\alpha}h_{\alpha}{}^{\beta}$, $\Phi_{\beta}{}^{\alpha} = \lambda^{\delta}\Omega_{\delta\,\beta}{}^{\alpha}$, and $\mathcal{T}_{a}{}^{\beta} = \lambda^{\alpha}T_{a\,\alpha}{}^{\beta}$. One then defines ghost number -2 operators acting on the ghost number three vertex operator as follows
\begin{eqnarray}
\hat{\Phi}^{a}U^{(3)} &=& \Phi^{a} +  Q(f^{a}) + (\lambda\Gamma^{a})_{\alpha}O^{\alpha}\label{phihata}\\
\hat{\Phi}^{\alpha}U^{(3)} &=& \Phi^{\alpha} - Q(O^{\alpha}) + \lambda^{\alpha}O_{\alpha}{}^{\beta} + \lambda^{\alpha}O\label{phihatalpha}\\
\hat{\Phi}_{\alpha}{}^{\beta}U^{(3)} &=& \Phi_{\alpha}{}^{\beta} + Q(O_{\alpha}{}^{\beta}) + Q(O) + (\lambda\Gamma^{a})_{\alpha}\mathcal{F}_{a}{}^{\beta}\label{phihatalphabeta}\\
\vdots\nonumber
\end{eqnarray}
where $f^{a}$, $O^{\alpha}$, $O_{\alpha}{}^{\beta} = (\Gamma^{ab})_{\alpha}{}^{\beta}O_{ab}$, $O$, $\mathcal{F}_{a}{}^{\alpha}$ are ghost number zero superfields to be determined. The non-BRST exact terms appearing on the right-hand side of eqns. \eqref{phihata}-\eqref{phihatalphabeta} are called shift-symmetry terms, and their origin lies in the dynamical structure of the model in study. The ellipsis below eqn. \eqref{phihatalphabeta} stands for further physical operators which will be irrelevant for our purposes. The 11D supergravity equations of motion then require these operators obey the relations
\begin{eqnarray}
\left[Q, \hat{\Phi}^{a}\right] + (\lambda\Gamma^{a})_{\alpha}\hat{\Phi}^{\alpha} &=& 0\label{Qphihata}\\
\{Q, \hat{\Phi}^{\alpha}\} - \lambda^{\beta}\hat{\Phi}_{\beta}{}^{\alpha} &=& 0\label{Qphihatalpha}\\
\left[Q , \hat{\Phi}_{\alpha}{}^{\beta}\right] + (\lambda\Gamma^{a})_{\alpha}\hat{\mathcal{T}}_{a}{}^{\beta} &=& 0\label{Qphihatalphabeta}
\end{eqnarray}
Notice that eqns. \eqref{phihata}-\eqref{phihatalphabeta} are consistent with \eqref{Qphihata}-\eqref{Qphihatalphabeta}.

\vspace{2mm}
In \cite{Cederwall:2009ez}, an explicit expression for $\hat{\Phi}^{a}$ was given. Its form can be cast as
\begin{eqnarray}\label{Ra}
\hat{\Phi}^{a} &=& -8\bigg[\frac{1}{\eta}(\bar{\lambda}\Gamma^{ab}\bar{\lambda})\partial_{b} - \frac{2}{\eta^{2}}(\bar{\lambda}\Gamma^{ab}\bar{\lambda})(\bar{\lambda}\Gamma^{cd}r)(\lambda\Gamma_{bcd}D)\nonumber\\
&& + \frac{32}{\eta^{3}}(\bar{\lambda}\Gamma^{ab}\bar{\lambda})(\bar{\lambda}\Gamma^{cd}r)(\bar{\lambda}\Gamma^{ef}r)(\lambda\Gamma_{e[f}\lambda)(\lambda\Gamma_{bcd]}w)\bigg]
\end{eqnarray} 
where $\eta = (\lambda\Gamma^{ab}\lambda)(\bar{\lambda}\Gamma_{ab}\bar{\lambda})$. After some algebraic manipulations, one shows that \eqref{Ra} indeed satisfies eqn. \eqref{Qphihata} with $\hat{\Phi}^{\alpha}$ given by
\begin{eqnarray}\label{defPhihatalpha}
\hat{\Phi}^{\alpha} &=& -\frac{32}{\eta^{2}}\xi^{\alpha}_{c}(\bar{\lambda}\Gamma^{cd}r)\partial_{d} + \frac{64}{\eta^{3}}\xi^{\alpha}_{e}(\bar{\lambda}\Gamma^{ef}r)(\bar{\lambda}\Gamma^{cd}r)(\lambda\Gamma_{fcd}D)\nonumber\\
&& + 64\left[Q, \frac{\xi^{\alpha}_{e}}{\eta^{3}}\right](\bar{\lambda}\Gamma^{ef}r)(\bar{\lambda}\Gamma^{cd}r)(\lambda\Gamma_{fcd}w)
\end{eqnarray}
where $\xi_c^\alpha$ takes the form
\begin{eqnarray}\label{defxicalpha}
\xi_c^\alpha &=& -2(\bar{\lambda}\Gamma_c)^\alpha (\bar\lambda \lambda) - 2(\bar{\lambda}\Gamma^b)^\alpha (\bar{\lambda}\Gamma_b \Gamma_c \lambda) + 2(\bar{\lambda}\Gamma_{bc})^\alpha (\lambda\Gamma^b \bar{\lambda}) + 4\bar{\lambda}^\alpha (\lambda\Gamma_c \bar\lambda)
\end{eqnarray}
One then let \eqref{Ra} act on the ghost number three wavefunction to find \cite{Berkovits:2018gbq}
\begin{eqnarray}\label{PhihataonU3}
\hat{\Phi}^{a}U^{(3)} &=& \Phi^{a} + Q\bigg[ \frac{48}{\eta^{2}}(\bar{\lambda}\Gamma^{ab}\bar{\lambda})(\bar{\lambda}\Gamma^{cd}r)(\lambda\Gamma_{bcd})^{\alpha}C_{\alpha\beta\delta}\lambda^{\beta}\lambda^{\delta}\nonumber\\
&& - \frac{24}{\eta}(\bar{\lambda}\Gamma^{ab}\bar{\lambda})C_{b\alpha\beta}\lambda^{\alpha}\lambda^{\beta}\bigg] + (\lambda\Gamma^{a}O)
\end{eqnarray}
with $O^{\alpha}$ being defined by
\begin{eqnarray}
O^{\alpha} 
&=& \frac{2}{\eta}\xi^{\alpha}_{c}\Phi^{c} - \frac{96}{\eta^{2}}\xi^{\alpha}_{e}(\bar{\lambda}\Gamma^{ed}r)C_{d\beta\delta}\lambda^{\beta}\lambda^{\delta} + \frac{192}{\eta^{3}}\xi^{\alpha}_{e}(\bar{\lambda}\Gamma^{ef}r)(\bar{\lambda}\Gamma^{cd}r)(\lambda\Gamma_{fcd})^{\alpha}C_{\alpha\beta\delta}\lambda^{\beta}\lambda^{\delta}\nonumber\\
\end{eqnarray}
Notice that \eqref{PhihataonU3} is consistent with \eqref{phihata}.

\vspace{2mm}
Likewise, one can compute the action of $\hat{\Phi}^{\alpha}$ on $U^{(3)}$. To this end, it will be useful to express \eqref{defxicalpha} in the more convenient way
\begin{eqnarray}
\xi^{\alpha}_{c} &=& -\frac{1}{2}(\Gamma_{abc}\lambda)^{\alpha}(\bar{\lambda}\Gamma^{ab}\bar{\lambda})
\end{eqnarray}
which follows from the 11D identity
\begin{eqnarray}\label{11didentity}
-(\Gamma^{ab})_{\alpha\beta}(\Gamma_{ab})_{\delta\epsilon} &=& 2(\Gamma^{a})_{\alpha\beta}(\Gamma_{a})_{\delta\epsilon} + 8(\Gamma^{a})_{\alpha(\delta}(\Gamma_{a})_{\epsilon)\beta} - 8C_{\alpha(\delta}C_{\epsilon)\beta}
\end{eqnarray}
where $C_{\alpha\beta}$ is the standard antisymmetric 11D spinor metric. One can check this identity by contracting with various p-form gamma matrices on both sides of \eqref{11didentity}. Furthermore, one can show that \eqref{11didentity} reduces to the standard 10D identity
\begin{eqnarray*}
(\gamma^{mn})^{\mu}{}_{\rho}(\gamma_{mn})^{\nu}{}_{\sigma} &=& 4(\gamma^{m})^{\mu\nu}(\gamma_{m})_{\rho\sigma} - 2\delta^{\mu}_{\rho}\delta^{\nu}_{\sigma} - 8\delta^{\mu}_{\sigma}\delta^{\nu}_{\rho}
\end{eqnarray*}
after dimensional reduction. Then, one obtains
\begin{eqnarray}
\hat{\Phi}^{\alpha}U^{(3)} 
&=& -32\bigg[\bigg\{Q,\frac{6}{\eta^{3}}\xi^{\alpha}_{e}(\bar{\lambda}\Gamma^{ef}r)(\bar{\lambda}\Gamma^{cd}r)(\lambda\Gamma_{fcd})^{\beta}C_{\beta\delta\epsilon}\lambda^{\delta}\lambda^{\epsilon} - \frac{3}{\eta^{2}}\xi^{\alpha}_{e}(\bar{\lambda}\Gamma^{ed}r)C_{d\beta\delta}\lambda^{\beta}\lambda^{\delta}\bigg\}\nonumber\\
&& + \big\{Q,\frac{1}{16}\frac{\xi^{\alpha}_{c}}{\eta}\Phi^{c}\big\} + \frac{1}{16}\frac{\xi^{\alpha}_{c}}{\eta}(\lambda\Gamma^{c})_{\alpha}\Phi^{\alpha} + \frac{1}{8\eta}\big[Q,\frac{\xi^{\alpha}_{c}}{\eta}\big](\lambda\Gamma^{c}\xi_{f})\Phi^{f}\nonumber\\
&& + \frac{3}{\eta^{2}}\big[Q, \xi^{\alpha}_{e}\big](\bar{\lambda}\Gamma^{ed}r)C_{d\beta\delta}\lambda^{\beta}\lambda^{\delta} - \frac{6}{\eta^{3}}\xi^{\alpha}_{e}(\lambda\Gamma^{e}\big[Q,\xi_{d}\big])(\bar{\lambda}\Gamma^{cd}r)C_{c\beta\delta}\lambda^{\beta}\lambda^{\beta}\bigg]\label{Phihatalphau3}
%
\end{eqnarray}
After making use of the identity
\begin{eqnarray}\label{identityforxialphac}
(\lambda\Gamma_{c})_{\alpha}\xi^{c\,\beta} &=& -\frac{1}{2}\delta^{\beta}_{\alpha}\eta - \frac{1}{4}(\Gamma^{c}\Gamma^{a})^{\beta}{}_{\alpha}(\lambda\Gamma^{b}{}_{c}\lambda)(\bar{\lambda}\Gamma_{ab}\bar{\lambda}) + \frac{1}{2}(\lambda\Gamma^{cb})_{\alpha}(\lambda\Gamma_{c}{}^{a})^{\beta}(\bar{\lambda}\Gamma_{ab}\bar{\lambda}) \nonumber\\
&& - \frac{1}{2}(\lambda\Gamma^{ab})_{\alpha}\lambda^{\beta}(\bar{\lambda}\Gamma_{ab}\bar{\lambda}) - \frac{1}{2}(\lambda\Gamma^{a})_{\alpha}(\lambda\Gamma^{b})^{\beta}(\bar{\lambda}\Gamma_{ab}\bar{\lambda})
\end{eqnarray}
eqn. \eqref{Phihatalphau3} becomes
\begin{eqnarray}
\hat{\Phi}^{\alpha}U^{(3)} &=& \Phi^{\alpha} + \bigg\{Q,-2\frac{\xi^{\alpha}_{c}}{\eta}\Phi^{c} + \frac{96}{\eta^{2}}\xi^{\alpha}_{e}(\bar{\lambda}\Gamma^{ed}r)C_{d\beta\delta}\lambda^{\beta}\lambda^{\delta}\nonumber\\
&& - \frac{192}{\eta^{3}}\xi^{\alpha}_{e}(\bar{\lambda}\Gamma^{ef}r)(\bar{\lambda}\Gamma^{cd}r)(\lambda\Gamma_{fcd})^{\beta}C_{\beta\delta\epsilon}\lambda^{\delta}\lambda^{\epsilon}\bigg\} + \lambda^{\alpha}O + (\lambda\Gamma^{ab})^{\alpha}O_{ab}\label{finalhatPhialphau3}
\end{eqnarray}
where $O$ and $O_{ab}$ can easily be determined from eqns. \eqref{Phihatalphau3}, \eqref{identityforxialphac}. Notice that \eqref{finalhatPhialphau3} is consistent with \eqref{phihatalpha}.

\vspace{2mm}
One can continue this procedure and define the remaining physical operators. For instance, the action of Q on $\hat{\Phi}^{\alpha}$ in \eqref{defPhihatalpha} reads
\begin{eqnarray}
\left\{Q, \hat{\Phi}^{\alpha}\right\} 
&=&\frac{64}{\eta^{3}}\phi\,\xi^{\alpha}_{c}(\bar{\lambda}\Gamma^{cd}r)\partial_{d} + \frac{64}{\eta^{3}}(\Gamma^{abe}\lambda)^{\alpha}(\lambda\Gamma_{e}\xi_{d})(\bar{\lambda}\Gamma_{ab}r)(\bar{\lambda}\Gamma^{cd}r)\partial_{c}\label{qphialpha}
\end{eqnarray}
where $\phi = (\lambda\Gamma^{ab}\lambda)(\bar{\lambda}\Gamma_{ab}r)$. It is not hard to see that \eqref{qphialpha} is consistent with \eqref{Qphihata}. In this way, the operator $\hat{\Phi}_{\alpha}{}^{\beta}$ can be extracted from \eqref{qphialpha}. We will omit the details of these computations and take $\hat{\Phi}_{\alpha}{}^{\beta}$ as being an actual operator.

\vspace{2mm}
Therefore, the operator $\hat{\Sigma}$ defined to be
\begin{eqnarray}\label{Sigmahat}
\hat{\Sigma} &=& P_{a}\hat{\Phi}^{a} + d_{\alpha}\hat{\Phi}^{\alpha} - \lambda^{\alpha}w_{\beta}\hat{\Phi}_{\alpha}{}^{\beta}
\end{eqnarray}
will map the ghost number three vertex $U^{(3)}$ onto a ghost number one pure spinor superfield. One can readily see from eqns. \eqref{phihata}-\eqref{phihatalphabeta} this ghost number one object is nothing but the ghost number one vertex operator \eqref{definitionofU1} up to a BRST-exact term which can be ignored since it will decouple from any physical observable. Hence, one learns that
\begin{eqnarray}\label{Sigmahatmap}
U^{(1)} &=& \left[\hat{\Sigma}, U^{(3)}\right]
\end{eqnarray}
This relationship can be viewed as a generalization of the ten-dimensional one relating the ghost number one and zero vertices through the b-ghost expressed in terms of descending operators \cite{Chang:2014nwa}. Since this last relation allows one easily to show the cyclic symmetry property of tree amplitudes in ordinary string theory correlators, one would expect \eqref{Sigmahatmap} to play a similar role in the 11D scattering amplitude prescription \eqref{11amplitudeprescription}. We will come back to this point later on after discussing the construction of the ghost number zero and two vertex operators.

\section{Ghost number zero vertex operator}\label{section4}
The 10D ghost number one and zero vertex operators are related to each other through either the non-minimal pure spinor or Y-formalism b-ghost \cite{Oda:2007ak} as shown in \cite{Grassi:2009fe}. We briefly review this 10D property in Appendix \ref{appendixA}. Here we use the non-Lorentz covariant or Y-formalism-like worldline 11D b-ghost given by
\begin{eqnarray}\label{bghost}
b &=& \frac{Y_{\alpha}G^{\alpha}}{2}
\end{eqnarray}
where $Y_{\alpha}= \frac{\nu_{\alpha}}{\lambda\nu}$, $G^{\alpha} = (\Gamma^{a}d)^{\alpha}P_{a}$, and $\nu_{\alpha}$ is a fixed pure spinor. It is easy to see that $\{Q, b\} = -\frac{P^{2}}{2}$, and its Lorentz-transformation is BRST-exact. This last property straightforwardly follows from the relation $[Q, H^{[\alpha\beta]}] = \lambda^{[\alpha}G^{\beta]}$, where $H^{[\alpha\beta]}$ is the bispinor appearing in the construction of the non-minimal 11D b-ghost \cite{Cederwall:2012es,Berkovits:2017xst}
\begin{eqnarray}\label{nonminimalbghost}
b_{n.m} &=& \frac{1}{2\eta}(\bar{\lambda}\Gamma^{ab}\bar{\lambda})(\lambda\Gamma^{ab}G) + \frac{1}{\eta^{2}}(\bar{\lambda}\Gamma^{ab}\bar{\lambda})(\bar{\lambda}\Gamma^{cd}r)(\lambda\Gamma^{ab})_{\alpha}(\lambda\Gamma^{cd})_{\beta}H^{[\alpha\beta]} + \ldots
\end{eqnarray}
where $\eta = (\lambda\Gamma^{ab}\lambda)(\bar{\lambda}\Gamma_{ab}\bar{\lambda})$. Therefore, the variation of \eqref{bghost} under Lorentz transformations with gauge parameter $\Lambda^{\alpha}{}_{\beta}$ is simply given by
\begin{eqnarray}
\delta_{\Lambda} b &=& -\frac{1}{2}\lambda^{\alpha}\Lambda^{\beta}{}_{\alpha}Y_{\beta}Y_{\delta}G^{\delta} + \frac{1}{2}Y_{\alpha}\Lambda^{\alpha}{}_{\beta}G^{\beta} = [Q, \frac{1}{2}Y_{\alpha}\Lambda^{\alpha}{}_{\epsilon}Y_{\beta}H^{[\beta\epsilon]}]
\end{eqnarray}

\vspace{2mm}
One then defines the ghost number zero vertex operator as
\begin{eqnarray}\label{u0u1}
U^{(0)} &=& \{b,U^{(1)}\}
\end{eqnarray}
By construction, this operator will automatically satisfy the descent equation $\{Q, U^{(0)}\} = \partial_{\tau}U^{(1)}$. Explicitly, one finds that
\begin{eqnarray}\label{actionofbonU}
\{b, U^{(1)}\} &=& \frac{1}{2}\lambda^{\alpha}(Y\Gamma^{b})^{\beta}D_{\beta}h_{\alpha}{}^{a}P_{a}P_{b} - \frac{1}{2}\lambda^{\alpha}(Y\Gamma^{b}d)\partial_{b}h_{\alpha}{}^{a}P_{a} - \frac{1}{2}\lambda^{\alpha}(Y\Gamma^{b})^{\beta}D_{\beta}h_{\alpha}{}^{\delta}d_{\delta}P_{b}\nonumber\\
&& + \frac{1}{2}(Y\Gamma^{b}d)\lambda^{\alpha}\partial_{b}h_{\alpha}{}^{\beta}d_{\beta} - \frac{1}{4}(Y\Gamma^{a})^{\sigma}\lambda^{\alpha}D_{\sigma}\Omega_{\alpha\,bc}N^{bc}P_{a} + \frac{1}{4}(Y\Gamma^{a}d)\lambda^{\alpha}\partial_{a}\Omega_{\alpha\,bc}N^{bc}\nonumber\\
&& + \frac{1}{2}P^{2}\lambda^{\alpha}h_{\alpha}{}^{\beta}Y_{\beta} + \frac{1}{8}P^{a}(Y\Gamma_{a}d)\lambda^{\alpha}\Omega_{\alpha\,bc}(\lambda\Gamma^{bc}Y)
\end{eqnarray}
The use of eqns. \eqref{eomforU1a}, \eqref{eomforU1g} allows us to write the first term of \eqref{actionofbonU} in the form
\begin{eqnarray}\label{1st}
\frac{1}{2}\lambda^{\alpha}(Y\Gamma^{b})^{\beta}D_{\beta}h_{\alpha}{}^{a}P_{a}P_{b} &=& \bigg\{Q,-\frac{1}{2}(Y\Gamma^{b})^{\beta}h_{\beta}{}^{a}P_{a}P_{b} + \frac{1}{2}(d\Gamma^{b}Y)h_{ab}P^{a}\bigg\} + h_{ab}P^{a}P^{b}\nonumber\\
&& - \frac{1}{2}\lambda^{\alpha}(Y\Gamma^{b})^{\beta}h_{\beta}{}^{\delta}(\Gamma^{a})_{\delta\alpha}P_{a}P_{b} + \frac{1}{2}(d\Gamma^{b}Y)\lambda^{\alpha}\partial_{b} h_{\alpha}{}^{a}P_{a}\nonumber\\
&& + \frac{1}{2}(d\Gamma^{b}Y)\lambda^{\alpha}h_{b}{}^{\beta}(\Gamma^{a})_{\beta\alpha}P_{a} - \frac{1}{2}P^{2}\lambda^{\alpha}h_{\alpha}{}^{\beta}Y_{\beta}\nonumber\\
&& + \frac{1}{2}(d\Gamma^{b}Y)\lambda^{\alpha}\Omega_{\alpha\,b}{}^{a}P_{a}
\end{eqnarray}
Furthermore, eqns. \eqref{eomforU1b}, \eqref{eomforU1e} imply that the third term in \eqref{actionofbonU} can be cast as
\begin{eqnarray}\label{3rd}
- \frac{1}{2}\lambda^{\alpha}(Y\Gamma^{a})^{\beta}D_{\beta}h_{\alpha}{}^{\delta}d_{\delta}P_{a} &=& 
\bigg\{Q,\frac{1}{2}(Y\Gamma^{a})^{\beta}h_{\beta}{}^{\delta}d_{\delta}P_{a}  - \frac{1}{2}(Y\Gamma^{a}d)h_{a}{}^{\delta}d_{\delta}\bigg\} + \frac{1}{2}(Y\Gamma^{a})^{\beta}h_{\beta}{}^{\delta}(\lambda\Gamma^{b})_{\delta}P_{b}P_{a}\nonumber\\
&& - h_{a}{}^{\delta}d_{\delta}P^{a} - \frac{1}{8}(Y\Gamma^{c})^{\beta}\Omega_{\beta\,ab}(\lambda\Gamma^{ab}d)P_{c} - \frac{1}{2}\lambda^{\alpha}(Y\Gamma^{a})^{\beta}\Omega_{\alpha\beta}{}^{\delta}d_{\delta}P_{a}\nonumber\\
&& - \frac{1}{2}(Y\Gamma^{a}d)\lambda^{\alpha}\partial_{a}h_{\alpha}{}^{\delta}d_{\delta} - \frac{1}{2}(Y\Gamma^{a}d)\lambda^{\alpha}T_{a\,\alpha}{}^{\delta}d_{\delta} + \frac{1}{2}(Y\Gamma^{a}d)\lambda^{\alpha}\Omega_{a\,\alpha}{}^{\delta}d_{\delta}\nonumber\\
&& - \frac{1}{2}(Y\Gamma^{a}d)h_{a}{}^{\delta}(\Gamma^{b}\lambda)_{\delta}P_{b}
\end{eqnarray}
In addition, the fifth and sixth terms in \eqref{actionofbonU} become, respectively
\begin{eqnarray}\label{5th}
-\frac{1}{4}(Y\Gamma^{a})^{\sigma}\lambda^{\alpha}D_{\sigma}\Omega_{\alpha\,bc}N^{bc}P_{a} &=&  \bigg\{Q,\frac{1}{4}(Y\Gamma^{a})^{\sigma}\Omega_{\sigma\,bc}N^{bc}P_{a}\bigg\} + \frac{1}{8}(Y\Gamma^{a})^{\beta}\Omega_{\beta\,bc}(\lambda\Gamma^{bc}d)P_{a}\nonumber\\
&& - 
\frac{1}{2}(Y\Gamma^{a})^{\sigma}\lambda^{\alpha}R_{\sigma \alpha,\delta}{}^{\beta}\lambda^{\delta}w_{\beta}P_{a} - \frac{1}{2}(Y\Gamma^{b}\Gamma^{a}\lambda)P_{b}\Omega_{a\,\alpha}{}^{\beta}\lambda^{\alpha}w_{\beta}\label{5th}\\
\frac{1}{4}(Y\Gamma^{a}d)\lambda^{\alpha}\partial_{a}\Omega_{\alpha\,bc}N^{bc} &=& \bigg\{Q,-\frac{1}{4}(Y\Gamma^{a}d)\Omega_{a\,bc}N^{bc}\bigg\} - \frac{1}{2}P^{a}N^{bc}\Omega_{a\,bc} + \frac{1}{4}(Y\Gamma^{d}\Gamma^{a}\lambda)P_{d}\Omega_{a\,bc}N^{bc}\nonumber\\
&& + \frac{1}{4}(Y\Gamma^{a}d)\lambda^{\alpha}R_{a\alpha\,bc}N^{bc} - \frac{1}{8}(Y\Gamma^{a}d)\Omega_{a\,bc}(\lambda\Gamma^{bc}d)\label{6th}
\end{eqnarray}
as a consequence of eqns. \eqref{eomforU1c}, \eqref{eomforU1d}, \eqref{eomforU1f}. Lastly, the 7th and 8th terms in \eqref{actionofbonU} add up to give a BRST-exact term,
\begin{eqnarray}\label{7th}
\frac{1}{2}\lambda^{\alpha}h_{\alpha}{}^{\beta}Y_{\beta}P^{2} + \frac{1}{8}P_{a}(Y\Gamma^{a}d)\lambda^{\alpha}\Omega_{\alpha\,bc}(\lambda\Gamma^{bc}Y)
&=& \bigg\{Q,-\frac{1}{2}P_{a}(Y\Gamma^{a}d)\lambda^{\beta}h_{\beta}{}^{\delta}Y_{\delta}\bigg\}
\end{eqnarray}
After plugging \eqref{1st}, \eqref{3rd}, \eqref{5th}, \eqref{6th},\eqref{7th} into \eqref{actionofbonU}, one arrives at the result
\begin{eqnarray}
\{b, U^{(1)}\} &=& 
\bigg\{ Q,\frac{1}{2}(Y\Gamma^{a})^{\beta}h_{\beta}{}^{\delta}d_{\delta}P_{a}  - \frac{1}{2}(Y\Gamma^{a}d)h_{a}{}^{\delta}d_{\delta} - \frac{1}{2}(Y\Gamma^{b})^{\beta}h_{\beta}{}^{a}P_{a}P_{b} + \frac{1}{2}(d\Gamma^{b}Y)h_{ab}P^{a}\nonumber\\
&& + \frac{1}{4}(Y\Gamma^{a})^{\sigma}\Omega_{\sigma\,bc}N^{bc}P_{a} - \frac{1}{4}(Y\Gamma^{a}d)\Omega_{a\,bc}N^{bc} - \frac{1}{2}(Y\Gamma^{a}d)P_{a}h_{\alpha}{}^{\delta}\lambda^{\alpha}Y_{\delta}\bigg\}\nonumber\\
&& + h_{ab}P^{a}P^{b} - h_{a}{}^{\delta}d_{\delta}P^{a} - \frac{1}{2}P^{a}N^{bc}\Omega_{a\,bc}  - \frac{1}{2}(Y\Gamma^{a}d)\lambda^{\alpha}T_{a\alpha}{}^{\beta}d_{\beta} - \frac{1}{2}P^{2}\lambda^{\alpha}h_{\alpha}{}^{\beta}Y_{\beta}\nonumber\\
&& + \frac{1}{2}(Y\Gamma^{a}d)\lambda^{\alpha}\Omega_{\alpha\,a}{}^{b}P_{b} - \frac{1}{2}(Y\Gamma^{a})^{\sigma}\lambda^{\alpha}R_{\sigma \alpha,\delta}{}^{\epsilon}\lambda^{\delta}w_{\epsilon}P_{a} - \frac{1}{2}\lambda^{\alpha}(Y\Gamma^{a})^{\beta}\Omega_{\alpha\beta}{}^{\delta}d_{\delta}P_{a}\nonumber\\
&& + \frac{1}{4}(Y\Gamma^{a}d)\lambda^{\alpha}R_{a\alpha ,bc}N^{bc}\label{partialresult}
\end{eqnarray}
One can simplify this expression further through the use of the following equations of motion
\begin{eqnarray}
R_{(\alpha\beta,\delta)}{}^{\epsilon} + (\Gamma^{a})_{(\alpha\beta}T_{a\,\delta)}{}^{\epsilon} &=& 0\nonumber\\
R_{\alpha\beta,\delta}{}^{\epsilon} - 2D_{(\alpha}\Omega_{\beta)\,\delta}{}^{\epsilon} + (\Gamma^{a})_{\alpha\beta}\Omega_{a\,\delta}{}^{\epsilon}&=& 0\nonumber\\
R_{a(\alpha,\beta)}{}^{\delta} + D_{(\alpha}T_{a\,\beta)}{}^{\delta}&=& 0
\end{eqnarray}
In doing so, one should be careful to keep the gauge invariance under the pure spinor constraint of \eqref{partialresult} at each step. For such a purpose, one introduces the 11D pure spinor projector $K_{\alpha}{}^{\beta}$ defined as $(\lambda\Gamma^{a})_{\beta}K_{\alpha}{}^{\beta} = -(\lambda\Gamma^{a})_{\alpha}$, and satisfies the identities $\lambda^{\alpha}K_{\alpha}{}^{\beta} = (\lambda\Gamma^{ab})^{\alpha}K_{\alpha}{}^{\beta} = 0$. An explicit expression for it was given in \cite{Cederwall:2012es} and reads
\begin{eqnarray}
K_{\alpha}{}^{\beta} &=& - \frac{1}{4}(Y\Gamma_{a})^{\beta}(\lambda\Gamma^{a})_{\alpha} - \frac{1}{2\beta}(Y\Gamma_{a})^{\beta}(\lambda\Gamma^{ab}\lambda)(Y\Gamma_{cb}Y)(\lambda\Gamma^{c})_{\alpha}\nonumber\\
&& + \frac{3}{4\beta}(Y\Gamma_{[ab})^{\beta}(Y\Gamma_{cd]}Y)(\lambda\Gamma^{ab}\lambda)(\lambda\Gamma^{cd})_{\alpha}
\end{eqnarray}
where $\beta = (\lambda\Gamma^{ab}\lambda)(Y\Gamma_{ab}Y)$. Using this object one replaces $w_{\alpha}$ in \eqref{partialresult} by the gauge-invariant object $\tilde{w}_{\alpha} = (\delta_{\alpha}^{\beta} + K_{\alpha}{}^{\beta})w_{\beta}$. Then, \eqref{partialresult} becomes
\begin{eqnarray}\label{finalbU}
\{b, U^{(1)}\} &=& h_{ab}P^{a}P^{b} - h_{a}{}^{\delta}d_{\delta}P^{a} - \frac{1}{2}P^{a}N^{bc}\Omega_{a\,bc} + P^{a}T_{a\,\delta}{}^{\beta}\lambda^{\delta}\tilde{w}_{\beta} + \frac{1}{2}(Y\Gamma^{a}d)\lambda^{\alpha}\Omega_{\alpha\,a}{}^{b}P_{b}\nonumber\\
&& -\frac{1}{2}P^{2}\lambda^{\alpha}h_{\alpha}{}^{\delta}Y_{\delta} + \frac{1}{2}(Y\Gamma^{a}d)T_{a\,\beta}{}^{\delta}\lambda^{\beta}K_{\delta}{}^{\epsilon}d_{\epsilon} + \frac{1}{2}(Y\Gamma^{a})^{\beta}P_{a}\lambda^{\alpha}\Omega_{\alpha\,\beta}{}^{\delta}K_{\delta}{}^{\epsilon}d_{\epsilon} + \{Q, \Lambda\}\nonumber\\
\end{eqnarray}
where $\Lambda$ is defined by the relation
\begin{eqnarray}\label{Lambda}
\Lambda &=& \frac{1}{2}(Y\Gamma^{a})^{\beta}h_{\beta}{}^{\delta}d_{\delta}P_{a}  - \frac{1}{2}(Y\Gamma^{a}d)h_{a}{}^{\delta}d_{\delta} - \frac{1}{2}(Y\Gamma^{b})^{\beta}h_{\beta}{}^{a}P_{a}P_{b} + \frac{1}{2}(d\Gamma^{b}Y)h_{ab}P^{a}\nonumber\\
&& + \frac{1}{4}(Y\Gamma^{a})^{\delta}\Omega_{\delta\,bc}N^{bc}P_{a} - \frac{1}{4}(Y\Gamma^{a}d)\Omega_{a\,bc}N^{bc} + \frac{1}{2}(Y\Gamma^{a}d)T_{a\,\beta}{}^{\delta}\tilde{w}_{\delta} + \frac{1}{2}(Y\Gamma^{a})^{\beta}P_{a}\lambda^{\alpha}\Omega_{\alpha\,\beta}{}^{\delta}\tilde{w}_{\delta}\nonumber\\
&& - \frac{1}{2}(Y\Gamma^{a}d)P_{a}h_{\alpha}{}^{\delta}\lambda^{\alpha}Y_{\delta}
\end{eqnarray}
Threfore, the ghost number zero vertex operator reads
\begin{eqnarray}\label{definitionofVU0}
U^{(0)} &=& h_{ab}P^{a}P^{b} - h_{a}{}^{\delta}d_{\delta}P^{a} - \frac{1}{2}P^{a}N^{bc}\Omega_{a\,bc} + P^{a}T_{a\,\delta}{}^{\beta}\lambda^{\delta}\tilde{w}_{\beta} + \frac{1}{2}(Y\Gamma^{a}d)\lambda^{\alpha}\Omega_{\alpha\,a}{}^{b}P_{b}\nonumber\\
&& -\frac{1}{2}P^{2}\lambda^{\alpha}h_{\alpha}{}^{\delta}Y_{\delta} + \frac{1}{2}(Y\Gamma^{a}d)T_{a\,\beta}{}^{\delta}\lambda^{\beta}K_{\delta}{}^{\epsilon}d_{\epsilon} + \frac{1}{2}(Y\Gamma^{a})^{\beta}P_{a}\lambda^{\alpha}\Omega_{\alpha\,\beta}{}^{\delta}K_{\delta}{}^{\epsilon}d_{\epsilon} 
\end{eqnarray}
As a double check, one can show that \eqref{definitionofVU0} indeed satisfies the relation $[Q, U^{(0)}] = \partial_{\tau}U^{(1)}$, as should. Note that $U^{(0)}$ possesses nice features: It is gauge-invariant under the pure spinor constraint, it is Lorentz-invariant up to BRST-exact and total derivative terms, and its variation under local gauge transformations is BRST-exact. Moreover, although this vertex explicitly depends on $Y_{\alpha}$, the latter always appears accompannied by a pure spinor variable, and so the whole vertex does seem to present a regular behaviour as $\lambda^{\alpha} \rightarrow 0$. All these features make $U^{(0)}$ a good candidate for calculating amplitudes. However, subtleties might be found when using the prescription \eqref{11amplitudeprescription}, since it only takes into account pure spinor variables zero modes. It is interesting to see that this same procedure applied for the 10D case (see Appendix \ref{appendixA}) provides a ghost number zero vertex independent of $Y_{\alpha}$.

\section{Ghost number two vertex operator}\label{section5}
One can also construct a ghost number two vertex operator by letting the non-Lorentz covariant b-ghost act on the ghost number three vertex $U^{(3)}$. Concretely,
\begin{eqnarray}
\{b, U^{(3)}\} &=& \frac{1}{2}(Y\Gamma^{a})^{\epsilon}P_{a}\lambda^{\alpha}\lambda^{\beta}\lambda^{\delta}D_{\epsilon}C_{\alpha\beta\delta} - \frac{1}{2}(Y\Gamma^{a}d)\lambda^{\alpha}\lambda^{\beta}\lambda^{\delta}\partial_{a}C_{\alpha\beta\delta}
\end{eqnarray}
which can be put into the convenient form
\begin{eqnarray}
\{b,U^{(3)}\} &=& - 3 P^{a}\lambda^{\alpha}\lambda^{\beta}C_{a\alpha\beta} -3(\lambda\Gamma^{ab}Y)(\lambda\Gamma_{a}d)\Phi_{b} + \bigg\{Q,-\frac{3}{2}(Y\Gamma^{a})^{\epsilon}P_{a}\lambda^{\beta}\lambda^{\delta}C_{\epsilon\beta\delta} \nonumber\\
&& + \frac{3}{2}(Y\Gamma^{a}d)\lambda^{\alpha}\lambda^{\beta}C_{a\alpha\beta}\bigg\}\nonumber
\end{eqnarray}
The use of the equations of motion \eqref{Qphihata}-\eqref{Qphihatalphabeta}, and the identity \eqref{identityforxialphac}, then imply that
\begin{eqnarray}\label{vertextwo}
\{b,U^{(3)}\} &=& -3P^{a}C_{a} + 3(\lambda\Gamma^{a}d)\Phi_{a} - 3(\lambda\Gamma^{a}w)(\lambda\Gamma_{a}\Phi) +  \bigg\{Q,-\frac{3}{2}(Y\Gamma^{a})^{\epsilon}P_{a}\lambda^{\beta}\lambda^{\delta}C_{\epsilon\beta\delta} \nonumber\\
&& + \frac{3}{2}(Y\Gamma^{a}d)\lambda^{\alpha}\lambda^{\beta}C_{a\alpha\beta} + 6(\lambda\Gamma^{a}\xi_{b})(\lambda\Gamma_{a}w)\Phi^{b}\bigg\}
\end{eqnarray}
The deduction of this equation made use of the specific form of $Y_{\alpha} = \frac{(\nu\Gamma^{ab}\nu)(\lambda\Gamma_{ab})_{\alpha}}{(\nu\Gamma^{ab}\nu)(\lambda\Gamma_{ab}\lambda)}$, which is also a valid choice in 11D, as discussed in next section. Hence, one learns that the Lorentz-covariant ghost number two vertex is given by
\begin{eqnarray}
U^{(2)} &=& - 3 P^{a}\lambda^{\alpha}\lambda^{\beta}C_{a\alpha\beta} + 3(\lambda \Gamma^{a} d)\Phi_{a} - 3(\lambda\Gamma^{a}w)(\lambda\Gamma_{a}\Phi)
\end{eqnarray}
Notice that this vertex is independent of $Y_{\alpha}$ as compared to $U^{(0)}$ in eqn. \eqref{definitionofVU0}, it is invariant up to BRST-exact and surface terms under super-gauge transformations, and satisfies the descent equation $[Q, U^{(2)}] = \partial_{\tau}U^{(3)}$.

\section{Discussion}\label{section6}
We have found a non-minimal pure spinor operator which reproduces $U^{(1)}$ from $U^{(3)}$. As was pointed out in section \ref{section3}, this relation can be interpreted as an 11D generalization of the 10D standard one \cite{Chang:2014nwa}. It might be potentially used for proving consistency properties (e.g. cyclic symmetry) of the 11D correlator \eqref{11amplitudeprescription} once the relation between $\hat{\Sigma}$ and $U^{(0)}$ is better understood.

\vspace{2mm}
We have also been succesful in finding a ghost number zero vertex operator which satisfies a standard descent equation, and possesses invariance under gauge symmetries up to BRST-exact or surface terms. However, it presents an explicit functional dependence on $Y_{\alpha} = \frac{\nu_{\alpha}}{\lambda\nu}$, which does not vanish even if $\lambda^{\alpha}$ is promoted to be a fully pure spinor, namely satisfying the additional constraint $\lambda\Gamma^{ab}\lambda = 0$, as suggested in \cite{Berkovits:2019szu}. On the other hand, the ghost number zero vertex defined via \eqref{u0u1} explicitly depends on the choice of the b-ghost, and so it might be not unique if one other b-ghost non-BRST-equivalent to \eqref{bghost} is shown to exist. An argument in favour of this last possibility goes as follows. 
 The structure of the variable $Y_{\alpha}$ is closely related to the linear term in $d_{\alpha}$ of the non-minimal b-ghost \cite{Berkovits:2005bt}. In 10D, this term has the structure $\frac{1}{\lambda\bar{\lambda}}(\bar{\lambda}\gamma^{m}d)P_{m}$. Thus, the 10D Y-formalism b-ghost \cite{Oda:2007ak} is immediately obtained from the non-minimal one after eliminating the $r$-variables and fixing $\bar{\lambda}_{\alpha}$ to a constant pure spinor, namely $\nu_{\alpha}$. If one generalizes this idea to 11D, one finds from \eqref{nonminimalbghost} that $Y_{\alpha}$ must actually be identified to the somewhat more complicated expression $\frac{(\nu\Gamma^{ab}\nu)(\lambda\Gamma_{ab})_{\alpha}}{(\nu\Gamma^{ab}\nu)(\lambda\Gamma_{ab}\lambda)}$. The non-Lorentz covariant b-ghost thus constructed also obeys $\{Q, b\} = -\frac{P^{2}}{2}$, and the only potential change one will find when doing the same manipulations of section \ref{section4} will be proportional to $\Omega_{\alpha\,\beta}{}^{\delta}$ in \eqref{definitionofU1}. Nevertheless, there exists one other subtlety related to the equivalence of the non-minimal and Y-formalism 11D b-ghosts which also deserves attention. In 10D, the cohomology of both b-ghosts are equivalent to each other \cite{Oda:2007ak}. On the other hand, the non-minimal 11D b-ghost is only nilpotent up to a BRST-exact term \cite{Berkovits:2017xst}. Thus, it is still unclear if the nilpotent 11D b-ghosts constructed out of either $Y_{\alpha} = \frac{\nu_{\alpha}}{\lambda\nu}$ or $Y_{\alpha} = \frac{(\nu\Gamma^{ab}\nu)(\lambda\Gamma_{ab})_{\alpha}}{(\nu\Gamma^{ab}\nu)(\lambda\Gamma_{ab}\lambda)}$ can play the same role of the non-minimal b-ghost in the construction of vertex operators. We leave the analysis and exploration of these issues for future research.

\section{Acknowledgments}
I am grateful to Oliver Schlotterer and Renann Lipinski Jusinskas for useful discussions and reading the manuscript. This research was supported by the European Research Council under ERC-STG-804286 UNISCAMP.

\begin{appendices}
\section{Relation between the 10D vertex operators through the Y-formalism b-ghost}\label{appendixA}
The 10D pure spinor superparticle is described by the action \cite{Berkovits:2001rb}
\begin{eqnarray}\label{10dpssuperparticle}
S &=& \int\,d\tau\,\bigg(P_{m}\partial_{\tau}X^{m} + p_{\alpha}\partial_{\tau}\theta^{\alpha} + w_{\alpha}\partial_{\tau}\lambda^{\alpha}\bigg)
\end{eqnarray}
and the BRST operator
\begin{eqnarray}\label{10dpsBRSToperator}
Q &=& \lambda^{\alpha}d_{\alpha}
\end{eqnarray}
where $d_{\alpha} = p_{\alpha} - \frac{1}{2}(\gamma^{m}\theta)_{\alpha}P_{m}$ is the fermionic constraint of the 10D Brink-Schwarz superparticle \cite{Brink:1981nb}. In this appendix we use Latin/Greek letters from the middle/beginning of the alphabet to denote 10D vector/Majorana-Weyl-spinor indices. The worldline fields $X^{m}$, $\theta^{\alpha}$ in \eqref{10dpssuperparticle} are the standard 10D superspace coordinates, and $P_{m}$, $p_{\alpha}$ are their corresponding conjugate momenta. Moreover, $\lambda^{\alpha}$ is a pure spinor satisfying the relation $\lambda\gamma^{m}\lambda = 0$, and $w_{\alpha}$ is its conjugate momentum which is defined up to the gauge transformation $\delta w_{\alpha} = (\gamma^{m}\lambda)_{\alpha}r_{m}$ for some arbitrary parameter $r_{m}$. The 10D Pauli matrices are denoted by $(\gamma^{m})_{\alpha\beta}$, $(\gamma^{m})^{\alpha\beta}$ and satisfy $(\gamma^{(m})^{\alpha\delta}(\gamma^{n)})_{\delta\beta} = \eta^{mn}\delta^{\alpha}_{\beta}$.

\vspace{2mm}
The BRST-cohomology of $Q$ \eqref{10dpsBRSToperator} is described by the 10D super-Yang-Mills states in its BV description. Thus, the gauge ghost, physical fields, antifields and gauge antighost can be shown to appear at ghost number 0,1,2 and 3 sectors of the Hilbert space, respectively. We will focus here on the ghost number one state, namely $V^{(1)} = \lambda^{\alpha}A_{\alpha}$. One can readily see this vertex describes the 10D super-Yang-Mills physical states by letting $Q$ act on it. One then obtains the conditions
\begin{eqnarray}
2D_{(\alpha}A_{\beta)} &=& (\gamma^{m})_{\alpha\beta}A_{m}\label{sym1}\\
\delta A_{\alpha} &=& D_{\alpha}\Lambda\label{sym2}
\end{eqnarray}
for some arbitrary superfield $\Lambda$. These equations \eqref{sym1}, \eqref{sym2} are exactly the superspace equations of motion and gauge transformations of 10D super-Yang-Mills. For this reason, $A_{\alpha}$ is naturally identified to the lowest mass-dimension component of the 10D super-gauge-connection.

\vspace{2mm}
On the other hand, a ghost number zero vertex $V^{(0)}$ satisfying the relation 
\begin{eqnarray}
\left[Q, V^{(0)}\right] &=& P^{m}\partial_{m}V^{(1)}
\end{eqnarray}
can be construced out of the worldline fields and the 10D super-Yang-Mills superfields. Its explicit form reads
\begin{eqnarray}\label{10dv0}
V^{(0)} &=& P_{m}A^{m} + d_{\alpha}W^{\alpha} + \frac{1}{2}N^{mn}F_{mn}
\end{eqnarray} 

\vspace{2mm}
Next, we will obtain \eqref{10dv0} from  $V^{(1)} = \lambda^{\alpha}A_{\alpha}$ through the relation
\begin{eqnarray}
V^{(0)} &=& \left\{b, V^{(1)}\right\}
\end{eqnarray}
where b is the Y-formalism b-ghost given by \cite{Oda:2007ak}
\begin{eqnarray}
b &=& \frac{(Y\gamma^{m}d)P_{m}}{2}
\end{eqnarray}
and $Y_{\alpha} = \frac{\nu_{\alpha}}{\lambda\nu}$, with $\nu_{\alpha}$ being a 10D fixed pure spinor. Then one has,
\begin{eqnarray}
\{b, V^{(1)}\} &=& \frac{1}{2}\lambda^{\alpha}(Y\gamma^{m})^{\beta}D_{\beta}A_{\alpha}P_{m} - \frac{1}{2}(Y\gamma^{m}d)\lambda^{\alpha}\partial_{m}A_{\alpha}
\end{eqnarray}
Using eqn. \eqref{sym1}, one finds that
\begin{eqnarray}
\{b, V^{(1)}\} 
&=& \bigg\{ Q, -\frac{1}{2}(Y\gamma^{m}A)P_{m}  + \frac{1}{2}(Y\gamma^{s}d)A_{s}\bigg\}  + A^{m}P_{m}  + \frac{1}{2}(Y\gamma^{s}d)\lambda^{\alpha}D_{\alpha}A_{s}\nonumber\\
&& - \frac{1}{2}(Y\gamma^{m}d)\lambda^{\alpha}\partial_{m}A_{\alpha}\label{bV1-a}
\end{eqnarray}
One can simplify \eqref{bV1-a} further by making use of the equation of motion
\begin{eqnarray}
D_{\alpha}A_{m} - \partial_{m}A_{\alpha} &=& (\gamma_{m}\chi)_{\alpha}
\end{eqnarray}
and the 10D gamma-matrix identity $(\gamma^{mn})_{\alpha}{}^{\beta}(\gamma_{mn})_{\delta}{}^{\epsilon} = 4(\gamma^{m})_{\alpha\delta}(\gamma_{m})^{\beta\epsilon} - 2\delta_{\alpha}^{\beta}\delta_{\delta}^{\epsilon} - 8\delta_{\alpha}^{\epsilon}\delta_{\delta}^{\beta}$. In this way, one learns that
\begin{eqnarray}
\{b, V^{(1)}\} 
&=& \bigg\{Q, -\frac{1}{2}(Y\gamma^{m}A)P_{m} + \frac{1}{2}(Y\gamma^{s}d)A_{s} +\frac{1}{4}J(Y\chi) - \frac{1}{4}N^{mn}(Y\gamma_{mn}\chi)\bigg\} + P^{m}A_{m}  +  d_{\alpha}\chi^{\alpha}\nonumber\\
&& + \frac{1}{4}N^{mn}(Y\gamma_{mn})_{\delta}\lambda^{\alpha}D_{\alpha}\chi^{\delta} - \frac{1}{4}JY_{\delta}\lambda^{\alpha}D_{\alpha}\chi^{\delta}\nonumber\\
\end{eqnarray}
where $N^{mn} = \frac{1}{2}(\lambda\gamma^{mn}w)$ is the ghost Lorentz current. Finally, one uses the equation of motion
\begin{eqnarray}
D_{\alpha}\chi^{\beta} &=& -\frac{1}{4}(\gamma^{mn})_{\alpha}{}^{\beta}F_{mn}
\end{eqnarray}
to show that
\begin{eqnarray}
\{b, V^{(1)}\}  
&=& \bigg\{Q , -\frac{1}{4}N^{mn}(Y\gamma_{mn}\chi) + \frac{1}{4}J(Y\chi)-\frac{1}{2}(Y\gamma^{m}A)P_{m} + \frac{1}{2}(Y\gamma^{s}d)A_{s}\bigg\}\nonumber\\
&& + (P^{m}A_{m} + d_{\alpha}\chi^{\alpha} + \frac{1}{2}N^{mn}F_{mn})
\end{eqnarray}
and therefore,
\begin{eqnarray}\label{10dv0}
V^{(0)} &=& P^{m}A_{m} + d_{\alpha}\chi^{\alpha} + \frac{1}{2}N^{mn}F_{mn}
\end{eqnarray}


\end{appendices}


\providecommand{\href}[2]{#2}\begingroup\raggedright\endgroup

\end{document}